\documentclass[aps,preprintnumbers,prl,twocolumn,superscriptaddress]{revtex4}

\usepackage{epsfig,latexsym,cancel,amssymb,amsmath}
\usepackage{graphicx}
\usepackage{epstopdf}
\usepackage{hyperref, graphicx, slashed, xcolor, amsmath}

\allowdisplaybreaks

\definecolor{red}{rgb}{1,0,0}

\newcommand{\beq}{\begin{equation}}
\newcommand{\eeq}{\end{equation}}
\newcommand{\bea}{\begin{eqnarray}}
\newcommand{\eea}{\end{eqnarray}}

\begin{document}

\title{Microlensing effect of charged spherically symmetric wormhole}

\author{Lei-Hua Liu}
\email{liuleihua8899@hotmail.com}
\affiliation{Department of Physics, College of Physics,
	Mechanical and Electrical Engineering,
	Jishou University, Jishou 416000, China}
\author{Mian Zhu}
\email{mzhuan@connect.ust.hk}
\affiliation{Department of Physics, The Hong Kong University of Science and Technology, Clear Water
	Bay, Kowloon, Hong Kong, P.R. China}
\affiliation{Jockey Club Institute for Advanced Study, The Hong Kong University of Science and Technology, Clear Water Bay, Kowloon, Hong Kong, P.R. China}

\author{Wentao Luo}
\email{wtluo@ustc.edu.cn}
\author{Yi-Fu Cai}
\email{yifucai@ustc.edu.cn}
\affiliation{CAS Key Laboratory for Research in Galaxies and Cosmology, Department of Astronomy, University of Science and Technology of China, Hefei, Anhui 230026, China}
\affiliation{School of Astronomy and Space Science, University of Science and Technology of China, Hefei 230026, China}
\author{Yi Wang}
\email{phyw@ust.hk}
\affiliation{Department of Physics, The Hong Kong University of Science and Technology, Clear Water
	Bay, Kowloon, Hong Kong, P.R. China}
\affiliation{Jockey Club Institute for Advanced Study, The Hong Kong University of Science and Technology, Clear Water Bay, Kowloon, Hong Kong, P.R. China}

\begin{abstract}
	
	We systematically investigate the microlensing effect of charged spherically symmetric wormhole, where the light source is remote from the throat. Remarkably, there will be at most three images by considering the charge part. We study all situations including three images, two images, and one image, respectively. The numerical result shows that the range of total magnification is from $10^5$ to $10^{-2}$ depending on various metrics. In the case of three images, there will be two maximal values of magnification (a peak, and a gentle peak) when the contribution via mass is much less than that of charge. However, we cannot distinguish the case that forms three images or only one image as the total magnification is of order $10^5$. Finally, our theoretical investigation could shed new light on exploring the wormhole with the microlensing effect.

\end{abstract}

\maketitle

\bigskip

\section{I.~~Introduction}
\label{introduction}

Wormhole \cite{Einstein:1935tc,Morris:1988cz} is an important topic in gravitational physics, which acts as a tunnel connecting two far separated regions. In the framework of General Relativity, maintaining wormhole throat needs exotic matter that violates the null energy condition \cite{Hochberg:1998ii, Hochberg:1998ha}. Hence, wormhole objects, if existed, would behave as a macroscopic negative mass and provide unique signals in astrophysics. 

In view of that, the gravitational lensing effect of a wormhole-like object attracts worldwide attentions. Most works concentrate on the lensing effect in strong regime, including both generic wormhole \cite{Safonova:2002si,TejeiroS:2005ltc,Nandi:2006ds,Tsukamoto:2012xs,Kuhfittig:2013hva,Nakajima:2014nba,Shaikh:2017zfl,Asada:2017vxl,Jusufi:2018kmk,Shaikh:2018oul,Shaikh:2019itn,Javed:2019qyg,Shaikh:2019jfr,Godani:2021aub}, and specific wormhole models like Ellis wormhole \cite{Yoo:2013cia,Takahashi:2013jqa,Izumi:2013tya,Tsukamoto:2016zdu,Tsukamoto:2016qro} and rotational symmetric wormhole \cite{SedighehHashemi:2015vht,Jusufi:2017mav}. Based on these works, even the wormhole is proposed to be observed \cite{Dai:2019mse,Simonetti:2020ivl,Bambi:2021qfo}. On the other hand, the microlensing effect of wormhole is less explored. The generic property of microlensing of wormhole is studied in \cite{Safonova:2001vz,Bogdanov:2008zy,Kitamura:2013tya}, and mostly applied in the case of an Ellis wormhole \cite{Abe:2010ap,Toki:2011zu,Tsukamoto:2017hva}.

Notably, the construction of the analytic wormhole solutions is usually difficult. The simplest wormhole, Ellis wormhole, is constructed before the concept of traversable wormhole \cite{Ellis:1973yv}, while wormholes beyonds Ellis is hard to construct. Hence, the study of microlensing effect is mostly applied in the case of an Ellis wormhole  \cite{Goulart:2017iko,Huang:2019arj}. However, recently there is growing interests on wormholes with electromagnetic effect. For example, a wormhole solution within Standard Model is constructed with a charged massless fermion \cite{Maldacena:2020sxe}, and wormholes with non-trivial mass and electric charge is founded in \cite{Blazquez-Salcedo:2020czn,Huang:2019arj}. 

Equipped with the concrete constructions, it is then interesting to explore the microlensing effect in a charged wormhole. Our work starts from a simple observation: a point charge with electric charge $Q$ would behave similarly to a positive mass. In the weak-field approximation, the charge $Q$ would contribute a $Q^2/r^2$ term to the effective Newtonian potential, which has the same sign as a positive mass, but the scaling on the distance $r$ differs. Hence, we expect a wormhole with a negative mass and a non-trivial charge to enjoy both features. More specifically, the wormhole should behave like a positive mass in the remoter region, and like a negative mass in the closer region.

This paper is organized as follows. In section II, we review the metric with the negative mass and the electric charge. In section III, we derive the Newtonian potential of the metric and obtain its corresponding deflection angle. In section IV, we systemamtically investigate the total magninification with various images. 
In section V, we give our main conclusion and the outlook for the exploring the microlensing effect within the wormhole.

\section{II.~~The Metric}
\label{metric}
In this paper, we will focus on the charged sperically symmetric wormhole, whose generic metric can be written as  
\begin{equation}
	\begin{aligned}
		ds^2=-h c^2 dt^2+(\sigma^2 h)^{-1}dr^2+r^2d\Omega^2,	
		\label{ellis huang1}
	\end{aligned}
\end{equation}
where $\sigma = \sigma(r)$ carries the information of the wormhole throat. Denote $r_0$ as the throat's radius, from the definition of wormhole throat, we have $\sigma(r_0) = 0$ and $r$ ranges from $r_0$ to infinity. Also, the absence of horizon requires $h(r) > 0$ for $r \in \left( r_0, \infty \right)$.

Our work starts from a simple observation: for a large variety of charged spherical symmetric wormholes, the function $h(r)$ would behave as
\begin{equation}
	h=1+\frac{h_MM}{r}+\frac{h_Q Q_e^2}{r^2},
	\label{ellis huang}
\end{equation}
where $M$ and $Q_e$ are the mass and electric charge of the wormhole, respectively. One may comprehensively understand \eqref{ellis huang} by noticing that, for remote regions i.e. $\sigma(r) \to 1$, the wormhole behaves as a compact object with mass $M$ and electric charge $Q_e$. Hence we expect the metric to behave similar to a RN metric
\begin{equation}
	ds^2 = h(r) c^2dt^2 - \frac{dr^2}{h(r)} - r^2 d\Omega^2 ~,
\end{equation}
with
\begin{equation}
	h(r;M,Q_e) \equiv 1 - \frac{2G}{c^2} \frac{M}{r} + \frac{G}{4\pi \epsilon c^4} \frac{Q_e^2}{r^2} ~.
\end{equation}

 For example, the charged Ellis wormhole \cite{Huang:2019lsl}, a generalization of Bronnikov-Ellis wormhole \cite{Bronnikov:1973fh,Ellis:1973yv},
the metric is of the form \eqref{ellis huang1} with 
\begin{equation}
	h = 1 - \frac{\gamma_2 Q^2}{4r r_0} \sigma + \frac{\gamma_1 Q^2}{4r_0^2} ~,~ \sigma^2 = 1 - \frac{r_0^2}{r^2} ~.
\end{equation}
The metric comes from the Lagrangian
\begin{equation}
	\mathcal{L}=\sqrt{-g}\bigg(R-\frac{1}{2} g^{\mu\nu}\partial_\mu\phi_\mu\partial_\nu\phi_\nu-\frac{1}{4}Z^{-1}F^2\bigg),
	\label{total lag}
\end{equation}
with 
\begin{equation}
	Z=\gamma_1\cosh(\phi)-\sinh(\phi)\gamma_2,
	\label{z}
\end{equation}
where $\phi$ is an external scalar field and $\gamma_1$, $\gamma_2$ are parameters. The ADM mass and electric charge are 
\begin{equation}
	M = \pm \frac{\gamma_2 Q^2}{8r_0} ~,~ Q_e = \gamma_1 Q ~,
\end{equation}
so the function can be rewritten as
\begin{equation}
	h(r) \equiv h(r;M,Q_e) = 1 \mp \frac{2M}{r} \sigma + \frac{Q_e^2}{4r^2 \gamma_1} ~.
\end{equation}
For remote region with $\sigma \simeq 1$, the function $h(r)$ is exactly of the form \eqref{ellis huang}.

Another example comes from the model in \cite{Blazquez-Salcedo:2020czn}. The Einstein-Dirac-Maxwell Lagrangian permits a wormhole solution
\begin{equation}
	ds^2 = \left( 1 - \frac{M}{r} \right)^2 dt^2 - \frac{dr^2}{\left( 1 - \frac{r_0}{r} \right) \left( 1 - \frac{Q_e^2}{r_0 r} \right)} - r^2 d\Omega^2 ~,
\end{equation}
where the mass $M$, the electric charge $Q_e$ and throat radius $r_0$ are related by 
\begin{equation}
	\label{eq:MQerelation}
	M = \frac{2Q_e^2 r_0}{Q_e^2 + r_0^2} ~.
\end{equation}
We can hence write it in the following form
\begin{equation}
	ds^2 = \left( 1 - \frac{2M}{r} + \frac{M^2}{r^2} \right)dt^2 - \frac{dr^2}{\left( 1 - \frac{2Q_e^2}{Mr} + \frac{Q_e^2}{r^2} \right)} - r^2 d\Omega^2 ~.
	\label{eq:BSwormhole}
\end{equation}
We see \eqref{eq:BSwormhole} can be of the form \eqref{ellis huang}, with the help of \eqref{eq:MQerelation}.

We see that for a large variety of wormholes, the property at regions remote from the wormhole throat $r \gg r_0$ can be described in a unified framework with \eqref{ellis huang}. In contrast to other compact objects with a non-trivial electric charges like RN black holes, the wormholes can permit a negative mass. As we shall see in the followings, the inclusion of a negative mass and a non-trivial electric charge can have non-trivial phenomena: those wormholes can lead to at most three images in microlensing. 
 
 \section{III.~~Lensing effect} 
 \label{lensing effect}
\subsection{Newtional potential approximation}
As discussed in section II, we come to the microlensing effect when both the sources and the observer are remote from the wormhole throat. Hence, we can simply take $\sigma = 1$. Meanwhile, it is also requiring that $\frac{h_MM}{r}\ll 1$ and $\frac{h_QQ_e^2}{r^2}\ll 1$ which will be confirmed by the later investigations. Moreover, we shall focus on the case when $M < 0$, since the $M>0$ case is well-explored. In this case, the spherically charged metric \eqref{ellis huang1} with \eqref{ellis huang} will be approximated by:
\begin{equation}
ds^2 = \left(1 - \frac{2\Phi}{c^2} \right) c^2dt^2 - \left(1 + \frac{2\Phi}{c^2}\right) dr^2 -r^2d\Omega^2 ~,
\label{total metric1}
\end{equation}
and with the Newtonian potential is depicted by
\begin{equation}
\label{eq:Phi}
\Phi =-\frac{h_M M c^2}{2r}-\frac{h_QQ_e^2c^2}{2r^2} ~.
\end{equation}
In appendix B, we show the new potential is the same under the isoptropic coordinate, in which we even keep the first order (higher order of $\Phi(\rho)$ can be considered as the GR correction). 
 The potential \eqref{eq:Phi} behaves differently for positive and negative mass. Comprehensively, for wormholes with $h_M M/h_Q <0$, the Newtonian potential is always non-zero and the light ray will ``feel attractive/repulsive forces''. Now if the mass changes its sign, the light ray can feel ``attractive'' or ``repulsive force'' in different spacetime locations. Hence, we expect the negavie mass from wormholes can bring rich phenomenology.

\subsection{Deflection angle}
\begin{figure}[ht]
		\includegraphics[scale=0.38]{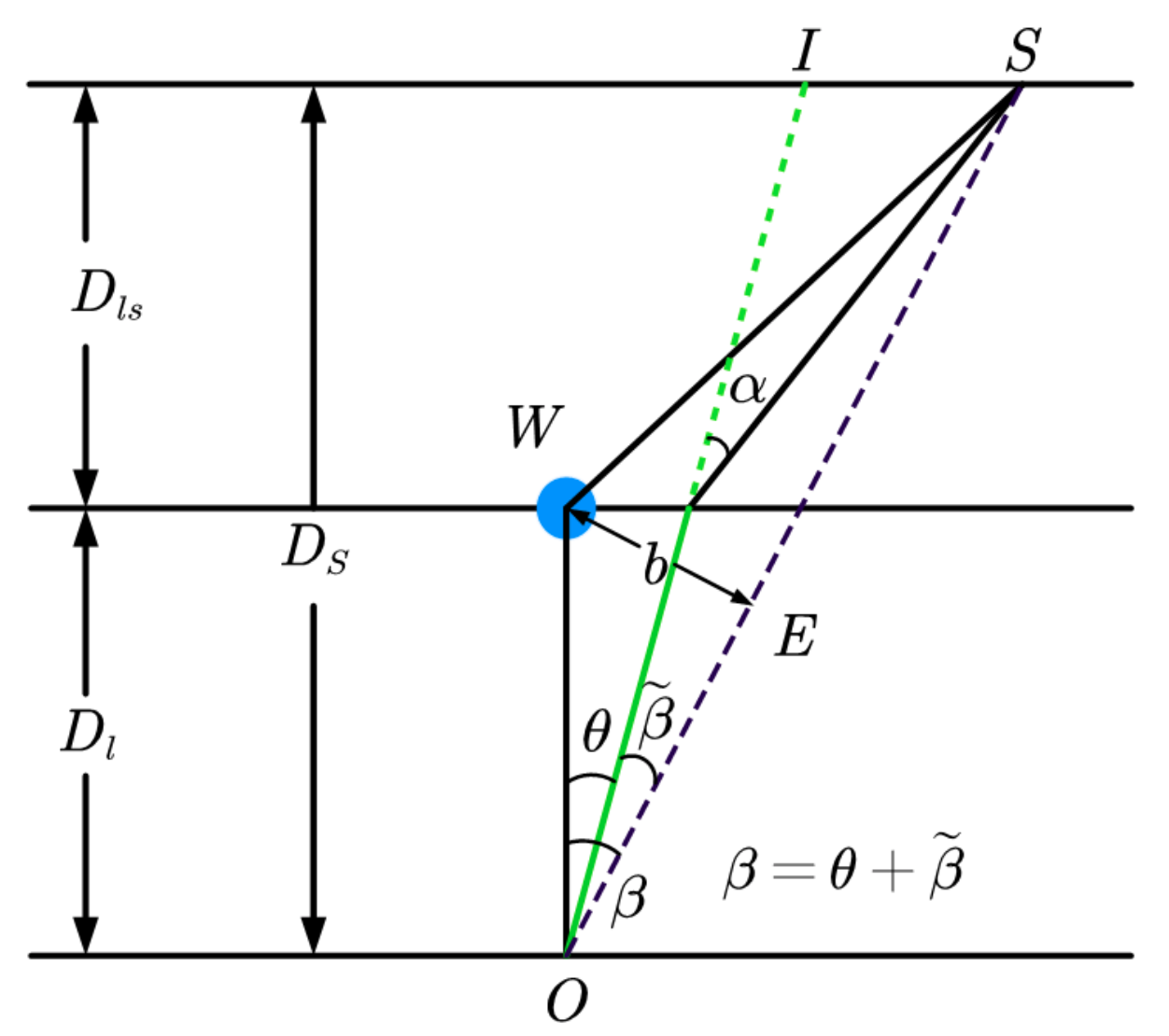}
		\caption{The lensing geometry via the wormhole source depicted by the blue ball $W$, in which the corresponding potential is \eqref{eq:Phi}. $O$ denotes the observer and $S$ is the source of light. The image is represented by $I$. $\alpha$ is the deflection angle and $\beta$ is the angle between the wormhole and light source. $\tilde{\beta}$ is the angle between the image and light source. $b$ is the impact parameter that is perpendicular to dotted line $OS$. $D_{ls}$, $D_l$ and $D_{s}$ are the angular diameter distances. For simplicity, we set $D_{ls}\approx D_{l}$. $z$ varies from $S$ to $O$.}
		\label{lensing}
\end{figure}

Before calculationg the deflection angle, the figure of lensing is shown in figure \ref{lensing}.

In light of this lensing geometry,  the distance of the photon from the wormhole can be expressed as $r=\sqrt{z^2+b^2}$ where $z$ denotes the distance along the unperturbed light ray from the point of closest approach, where it varies from $S$ to $O$ in figure \ref{lensing}. Then, the Newtonian potential can be represented by
\begin{equation}
\Phi = -\frac{c^2 h_M M}{2\sqrt{b^2 + z^2}} - \frac{h_Q c^2Q_e^2}{2(b^2+z^2)} ~.
\label{potental in b}
\end{equation}
One of the most essential ingredient in lensing is the deflection angle defined by the difference of the initial and final ray direction,
\begin{equation}
\label{eq:alphadef}
\vec{\alpha} = 2 \int \nabla_{\perp} \Phi dl ~, 
\end{equation}
where $\nabla_{\perp} \Phi$ is the projection of $\nabla \Phi$ onto the the plane orthogonal to the direction of the ray, in our case is the $\vec{b}$ direction, and hence
\begin{equation}
\nabla_{\perp} \Phi = \left[ \frac{h_M M c^2}{2(b^2 + z^2)^{\frac{3}{2}}} + \frac{h_Q c^2 Q_e^2}{(b^2 + z^2)^2} \right] \vec{b} ~.
\end{equation}
The integration of \eqref{eq:alphadef} takes place in $\vec{l}$ direction, in our case $dl \simeq dz$, so
\begin{equation}
\int \nabla_{\perp} \Phi dl = \vec{b} \left[ \frac{h_MMc^2z}{2b^2r} + \frac{h_Qc^2 Q_e^2}{2b^3} \left( \arcsin \frac{z}{r} + \frac{zb}{r^2} \right) \right] ~.
\end{equation}
Now we determine the range of integration. We are interested in the microlensing effect, where the impact parameter $b$ is much smaller than the distance of the source, lens, and observer. That is, $b \ll D_l$ and $b \ll D_{ls}$ (the requirement of microlensing). Translated into the integration, we have $z \simeq r \gg b$ in both the upper and lower bound of the integration, so
\begin{equation}
\label{eq:alpha}
\vec{\alpha} = \hat{b} \left[ \frac{2h_M M}{b} + \frac{\pi h_Q Q_e^2}{ b^2} + \mathcal{O}\left( \frac{b^2}{D_l^2} \right) + \mathcal{O}\left( \frac{b^2}{D_{ls}^2} \right) \right] ~,
\end{equation}
where $\hat{b}$ is the unit vector along the $\vec{b}$ direction, the integration range is from $-z$ to $z$ meaning that from $S$ (source of light) to $O$ (observer) and our calculation also implements the approximation $D_{ls}\approx D_l$. Due to the requirement of microlensing, we neglect the contribution of higher order of $\mathcal{O}\left( \frac{b^2}{D_l^2} \right) + \mathcal{O}\left( \frac{b^2}{D_{ls}^2} \right)$. Then, the lensing equation for angle $\beta$ can be represented by 
\begin{equation}
\vec{\beta} = \vec{\theta} - \frac{D_{ls}}{D_s} \vec{\alpha} ~,
\end{equation}
and we get the expression for $\beta$ in light of \eqref{eq:alpha}:
\begin{equation}
\label{eq:lens}
\beta = \theta -  \frac{D_{ls}}{D_sD_l \theta} \left( 2h_M M  + \frac{\pi h_Q  Q_e^2}{ b} \right) ~.
\end{equation}
Being armed with this essential quantity for the lensing effect, nextly we will proceed with the Einstein angle for exploring the images of light source. 

\subsection{Einstein radius}
\label{einstein radius}
Conventionally, the Einstein ring will be formed as the lens, source, and observer are perfected aligned, in which the lensing source can be considered as a particle-like positive mass object. Although our case is quite different from the traditional picture, the negative mass part and electric charge part can all be absorbed into the total Newtonian potential which is highly relevant to the Einstein angle. Thus, we still use a similar definition as Ref. \cite{Safonova:2001vz} that is defined by
\begin{equation}
\beta = \theta + \frac{\theta_E^2}{\theta} ~.
\label{beta in einstein angle}
\end{equation}
To be more precise, the difference in our case is from the dependence on impact parameter $b$. In light of figure \ref{lensing}, one can approximately obtain $b = D_l \theta$ (making use of $\sin(\theta)\approx \theta$ as $\theta\ll 1$), in which we have used the $\theta\approx \beta$ since the image of light will be quite near to the source $S$. After making these approximations, our lensing equations become
\begin{equation}
\label{eq:lensbeta}
\beta = \theta + \frac{\theta_{\rm E,M}^2}{\theta} - \frac{\theta_{\rm E,Q}^3}{\theta^2} ~\to~ \theta^3 - \beta \theta^2 + \theta_{\rm E,M}^2 \theta - \theta_{\rm E,Q}^3 = 0 ~,
\end{equation}
with 
\begin{equation}
\theta_{\rm E,M}^2 = - 2h_MM\frac{D_{ls}}{D_sD_l} ~,~ \theta_{E,Q}^3 = \frac{D_{ls} \pi h_QQ_e^2}{D_sD_l^2 } ~.
\label{main lens eq}
\end{equation}
Thus, this equation relates the massive paremeter $h_M$ and charge parameter $h_Q$. The classification of real solutions corresponds to the various images of the light source $S$, which means that the number of real solutions of eq. \eqref{main lens eq} corresponds to the number of images of $S$ after the $W$ (wormhole) bends the light. Its complete solution are written in Appendix VII. 

Then, we will briefly analyze the solutions for Eq. \eqref{main lens eq}. In Cases 1 and 2 of Appendix VII, there is only one real solution of Eq. \eqref{main lens eq}. Especially for Case $1$, $\theta_E=\frac{1}{3}\beta$ means that three images merges at the Einstein angular radius. One can find a detailed forming image in Ref. \cite{Safonova:2001vz}. As for Case $2$, it shows the image will be formed in a different place compared with Case $1$. These two cases all belong to $\Delta>0$ (whose definition can be found in Eq. \eqref{delta}). Secondly, as $\Delta=0$, one can find that there are two solutions of \eqref{main lens eq}, which leads to two images one is located inside the radius and the other one is outside the radius. The most striking feature is that there will be three images that is belonging to case 4 in Appendix VII. Then, we will particularly focus on this case for exploring its magnification. 

For observations, the most essential part is the magnification of images. In the following part, we will show the total magnification in various cases belonging to Appendix VII.

\section{IV.~~Magnification}
\label{magni}
 Similar to the optics, the magnification of images will occur as the light is bent by gravity which is caused by the wormhole. In this paper, we consider the light sources that are remote from the throat of the wormhole. Thus, the wormhole could be considered a point-like object. Following the standard procedure, the magnification can be defined by
 \begin{equation}
 \mu^{-1}_i = \left | \frac{\beta}{\theta} \frac{d\beta}{d\theta_i} \right | ~,
 \label{mag}
 \end{equation}
 where $\mu$ is the magnification and $\theta_i$ is the $i$th solution in the fourth case of Appendix VII. In this part, we will investigate the four cases of Appendix VII comparing with \cite{Tsukamoto:2017hva,Safonova:2001vz}. 
 Here, we have three magnifications $\mu_i$ corresponding to $\theta_i$ (the real solution of lens equation \pageref{sec:lenssol}). Meanwhile, we can simply derive the formula 
$\frac{\beta}{\theta}=1+\frac{\theta_{\rm E,M}^2}{\theta^2}-\frac{\theta_{\rm E,Q}^3}{\theta^3}$ and $\frac{d\beta}{d\theta}=1-\frac{\theta_{\rm E,M}^2}{\theta^2}+\frac{2\theta_{\rm E,Q}^3}{\theta^3}$. Thus, we can obtain the $i$th manification
\begin{equation}
	\mu_i=\bigg(1-\frac{\theta_{\rm E,M}^4}{\theta_i^4}+3\frac{\theta_{\rm E,M}^2\theta_{E,Q}^3}{\theta_i^5}+\frac{\theta_{\rm E,Q}^3}{\theta_i^3}-\frac{2\theta_{\rm EQ}^6}{\theta_i^6}\bigg)^{-1}.
	\label{magi}
\end{equation}
This formula is applicable for four cases in Appendix VII. From another aspect, the wormhole structure is absorbed in the $\theta_{\rm E, Q}$ and $\theta_{\rm E, M}$. Thus, our result can be naturally implemented into a travservable wormhole. Finally, the total magnification can be derived as
\begin{equation}
	\mu_{\rm total}=\sum_{i}|\mu_i|.
	\label{mutotal}
\end{equation}
where $i$ donotes the $i$th manification corresponding to Appendix VII. 
Being armed with this formula, one could investigate the four cases in Appendix \pageref{sec:lenssol}. 

\subsection{Three images}
\label{three images}
In this subsection, we will study the fourth case of Appendix VII, which could lead to three images after bending the light by a wormhole. 

Note that $\theta_{\rm E,Q}=0$, $\theta_{\rm E,M}$ are related to the structure of wormhole, and $b$ only depends on the location of light source, we could classcify the wormhole into three cases: (a) $\theta_{\rm E,M}\ll 1$, (b) $\theta_{\rm E,Q}\ll 1$ and (c) $\theta_{\rm E,M}$ is compatible with $\theta_{\rm E,Q}$. As $\theta_{\rm E,Q}=0$ and $\theta_{\rm E,M}=0$, it corresponds to the case that the wormhole is chargeless and massless, respectively. 

Before entering the detailed investigation, we need to figure out the range of these three parameters. Based on the requirement of microlensing, all of the angles including that the deflection angle need to be very small whose unit is around a few arcseconds. Consequently, $\theta_{\rm E,Q}$ and $\theta_{\rm E,M}$
are also small in light of Eq. \eqref{eq:lensbeta}, where we set the range of them as $0<\theta_{\rm E,Q}<1$ and $0<\theta_{\rm E,M}<1$ in natural units, respectively.

\subsubsection{Case a: $\theta_{\rm E,M}\ll \theta_{\rm E,Q}$}
\label{m=0}
In this case, we will take $\theta_{\rm E,M}=10^{-7}$ and vary $\beta$ and $\theta_{\rm E,Q}$. In the traditional procedure, people will use time to describe the magnification. 
In figure \ref{lensing}, it clearly shows the image of the light source. If the light source travels at a constant speed $v$, then $\tan(\theta)=\frac{S}{D_s}$ where $S=v t$ (the distance of travelling for $S$). In our previous calculation, we set $D_s=2D_l$, meanwhile combining with $\theta\ll 1$. Therefore, one can approximately obtain $S\approx b$, in which we obtain $\theta\approx \frac{b}{D_l}$ since $\theta\ll 1$. We restrict our range of the sacle within the Milky way whose size is around $10~\rm kpc$, thus the maximal value of $b$ is less than $10~\rm kpc$. For various light sources, their traveling speeds are different, thus we will use the impact parameter $b$ as a variable of total magnification $\mu_{\rm total}$. To be more precise, $b$ and $S$ are of the same order. 

  \begin{figure}[ht]
  	\includegraphics[scale=0.5]{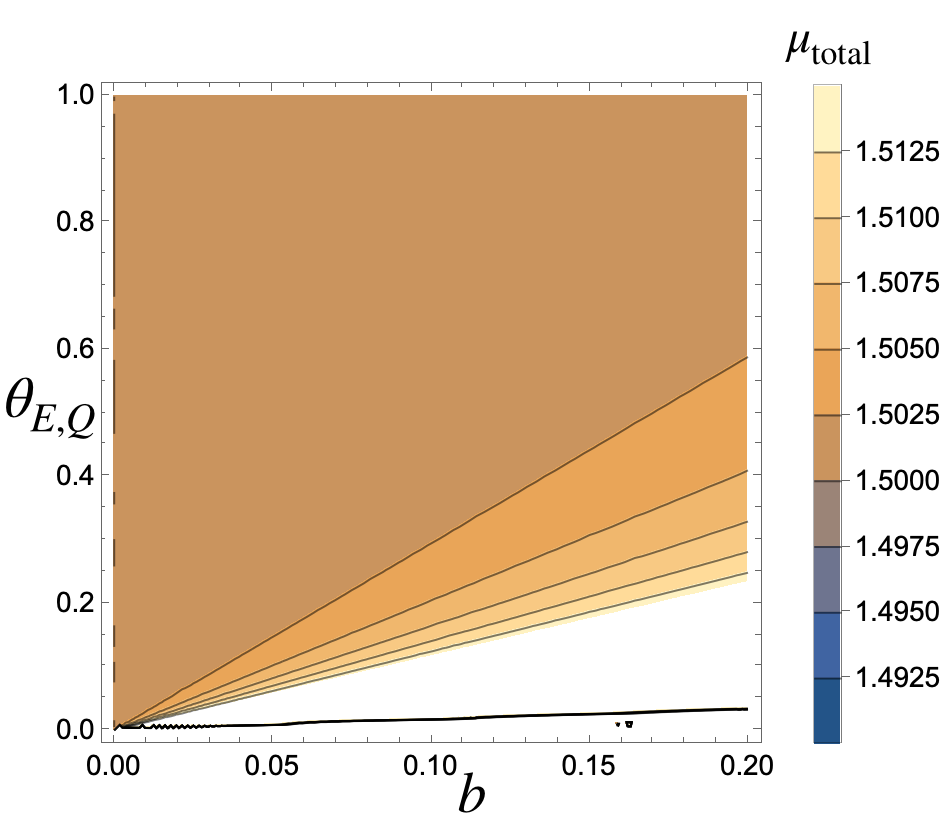}
  	\caption{The contour plot of $\mu_{\rm total}$ in terms of $\theta_{\rm E,Q}$ and $b$ as fixing $\theta_{\rm E,M}=10^{-7}$. We have set $0<b<0.2~\rm kpc$, $D_l=10~\rm kpc$, $0<\theta_{\rm E,Q}<1$. }
  	\label{mu11}
  \end{figure}
 
In figure \ref{mu11}, we plot $\mu_{\rm total }$ as a function of $b$ and $\theta_{\rm E,M}=10^{-7}$. The mass of wormhole is extremally small compared with the contribution from charge. It indicates that $\mu_{\rm total}$ will be approaching $1.5$ in most parameter space, even though we numerically obtain that $\mu_{\rm total}$ is almost one in large scales ($b$ approaches $10~\rm kpc$). Meanwhile, we could see that $\mu_{\rm total}=1.5$ can be dubbed as a critical value for assessing that the contribution comes via the electric charge part. 
As for the white part, it corresponds to $\mu_{\rm total}$ is of higher order even higher being of order of $10^4$), which is need to be investigated.

  \begin{figure}[ht]
  	\includegraphics[scale=0.42]{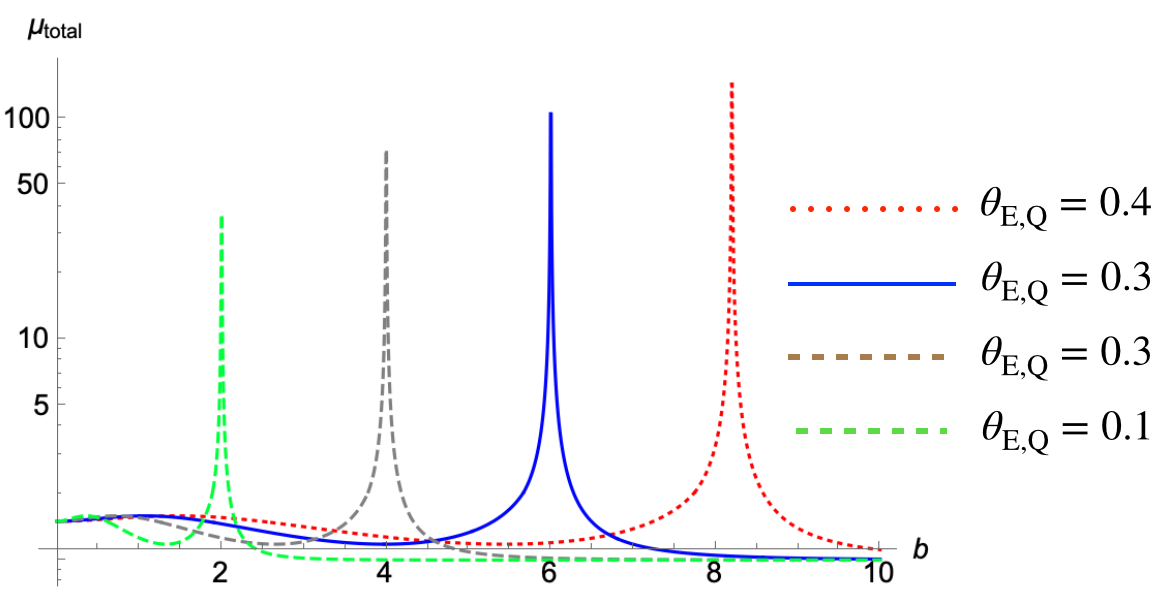}
  	\caption{The plot of $\mu_{\rm total}$ in terms of $\theta_{\rm E,Q}$ and $b$ as fixing $\theta_{\rm E,M}=10^{-7}$. We have set $0<b<10~ \rm kpc$, $D_l=10~\rm kpc$, $\theta_{\rm E,M}=10^{-7}$, $\theta_{\rm E,Q}=0.1$, $\theta_{\rm E,Q}=0.2$, $\theta_{\rm E,Q}=0.3$ and $\theta_{\rm E,Q}=.4$. }
  	\label{mu12}
  \end{figure}
From figure \ref{mu12}, one of our core results is that there will be some peaks and gentle peaks at different scales $b$, which gives an explanation for some observations at different time scales, in which the current observation for microlensing using the days. If there are two peaks within the observations can be comparable with our results, it will be of significance for exploring its origin. We also found that as $\theta_{\rm E,Q}>0.5$, $\mu_{\rm total}$ will be appoaching $1.5$ more or less.

\subsubsection{Case b: $\theta_{\rm E,Q}\ll \theta_{\rm E,M}$}
\label{q=0}
In this case, the contribution from mass part is much larger than charge part, where we set $\theta_{\rm E,Q}=10^{-4}$. Here, we adopt the total magnification \eqref{magi} for the numerical simulatation.  
 \begin{figure}[ht]
 	\includegraphics[scale=0.5]{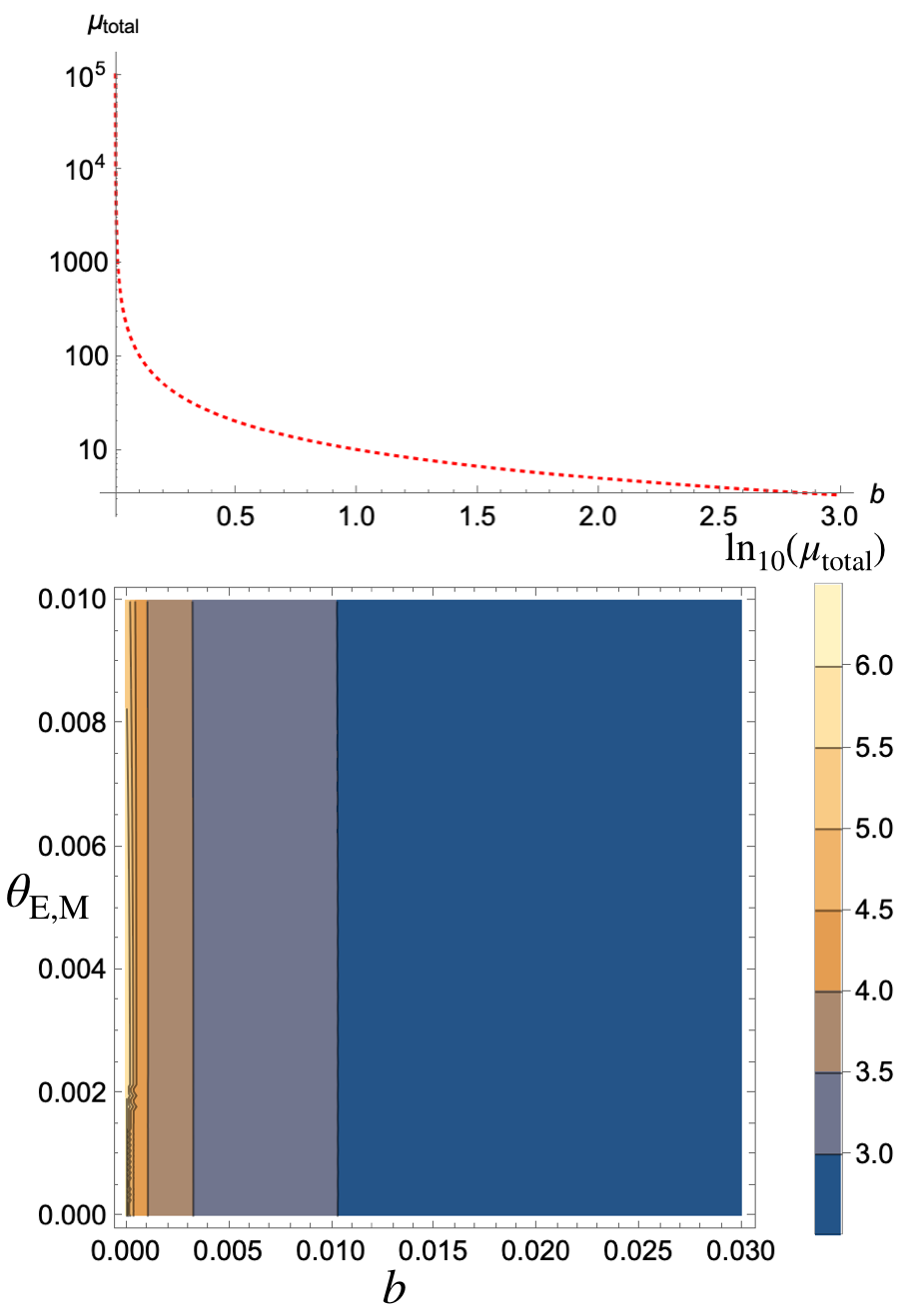}
 	\caption{The plot of $\mu_{\rm total}$ in terms of $\theta_{\rm E,M}$ and $b$. In the upper pannel, we fix $\theta_{\rm E,Q}=10^{-4}$ and $\theta_{\rm E,M}=0.8$. The lower pannel shows $\ln_{10}(\mu_{\rm total})$ in terms of $\theta_{\rm E,Q}$ and $b$ as fixing $\theta_{\rm E,Q}=10^{-4}$. We have set $0<b< 0.03~kpc$, $D_l=10~\rm kpc$.}
 	\label{mu21}
 \end{figure}
In figure \ref{mu21}, we have shown that $\mu_{total}$ is very large as $\theta_{\rm E,M}=10^{-4}$ and $\theta_{\rm E,M}=0.8$. When $\theta_{\rm E,M}>0.008$, $\mu_{\rm total}$ will be of order of $10^{5}$ and the trend of magnificaltion will lead to be the same (that is why we choose $\theta_{\rm E,M}=0.8$). And we also obtain that $\mu_{\rm total}< 1.5$ as choosing $b>1~\rm ~kpc$. In order to investigate the order of $\mu_{\rm total}$ as $b<0.03~\rm kpc$ ($\beta<10^{-3}$), we also numerically vary with respect to $\theta_{\rm E,M}$. 
 
 In the lower pannel of figure  \ref{mu21}, we show that $\ln_{10}(\mu_{\rm total})$ as a function of $b$ and $\theta_{\rm E,M}$, in which it clearly indicates that the order of $\mu_{\rm total}$ can reach to $10^{6}$ at the very small $b$, meanwhile, it also reveals that $ \mu_{\rm total}\approx 1$ as $\theta_{\rm E,M}$ will derease as impact parameter enhances.

 \subsubsection{Case c: Comparable $\theta_{\rm E,M}$ and $\theta_{\rm E,Q}$  }
\label{m=q}
In this case, we will consider the contributions from $\theta_{\rm E,M}$ and $\theta_{\rm E,Q} $ at the same time by setting $\theta_{\rm E,Q}$ and $\theta_{\rm E,M}$ are comparable, at least for which they are of the same order. 
 \begin{figure}[ht]
 	\includegraphics[scale=0.5]{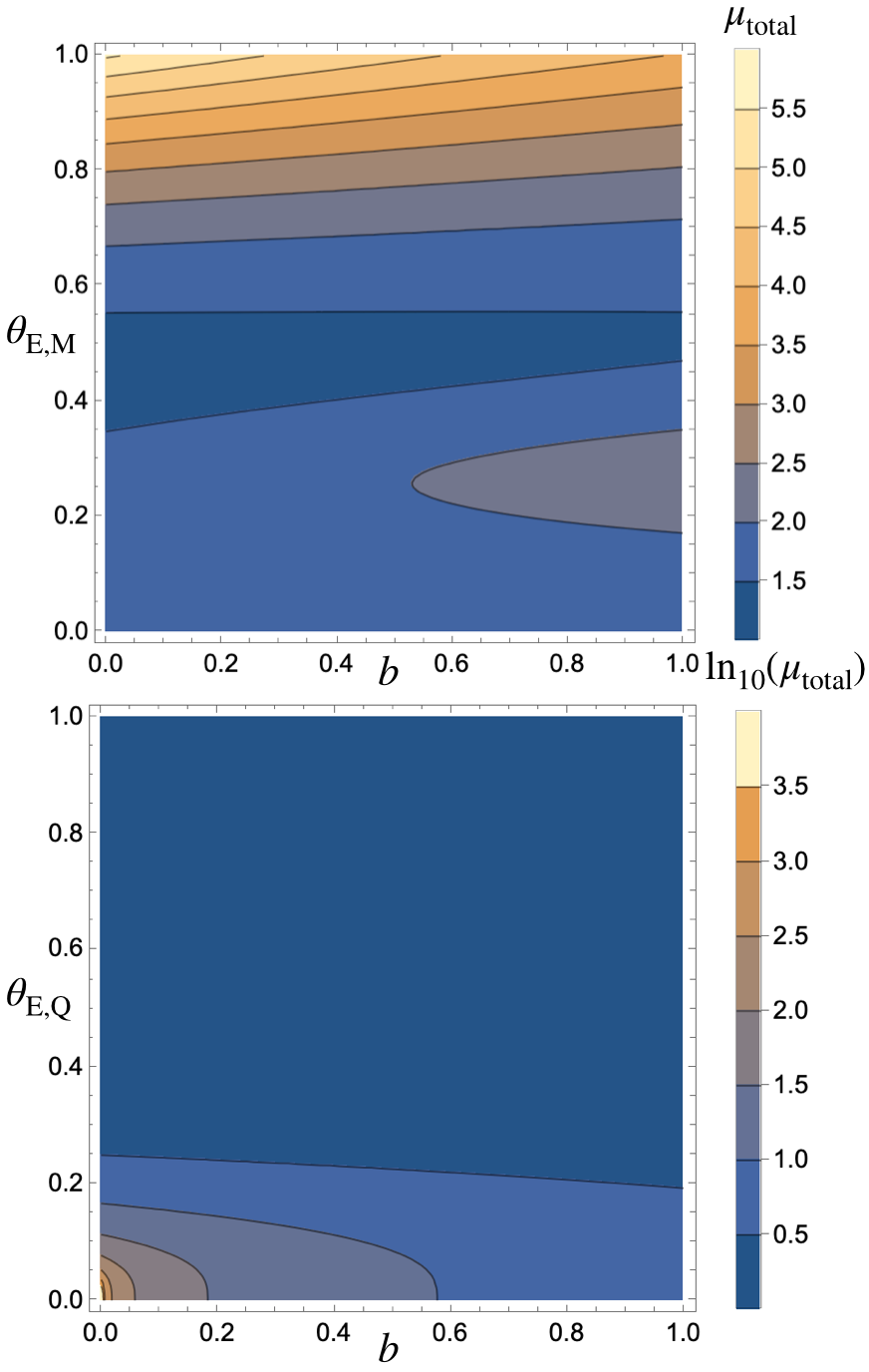}
 	\caption{The upper pannel is the contour plot of $\mu_{\rm total}$ in terms of $\theta_{E,M}$ and $b$ as fixing $\theta_{\rm E,Q}=0.45$. We have set $0<b< 1~\rm kpc$, $D_l=10~\rm kpc$. In the lower pannel is the contour plot of $\ln_{10}(\mu_{\rm total})$ in terms of $\theta_{\rm E,Q}$ and $b$ as fixing $\theta_{\rm E,M}=0.45$. We have set $0<b< 1~\rm kpc$, $D_l=10~\rm kpc$. }
 	\label{mu31}
 \end{figure}
In the upper pannel of figure \ref{mu31}, we plot $\mu_{\rm total}$, where we have set $\theta_{\rm E,Q}=0.45$ then vary with respect to $\theta_{\rm E,M}$ and $b$. The magnification is mainly affected by the $\theta_{\rm E,M}$, but not for $b$. Further, $\mu_{\rm total}$ is around $1.5$ even $b$ is large. The maximal value of $\mu_{\rm total}$ is around $6$. 

Subsequently, we will fix the value of $\theta_{E, M}$. In the lower pannel of figure \ref{mu31} we show that $\ln_{10}(\mu_{\rm total})$ as a function of $\theta_{E,Q}$ and $b$. Being different from figure \ref{mu31}, the $\mu_{\rm total}$ will be very large as varying with $\theta_{\rm E,Q}$. The most important information in the upper pannel of figure \ref{mu31} is that the order of magnification is from $10^{0.5}$ to $10^{3.5}$ which nearly covers the whole observational range for the magnification. 

\begin{figure}[ht]
 	\includegraphics[scale=0.46]{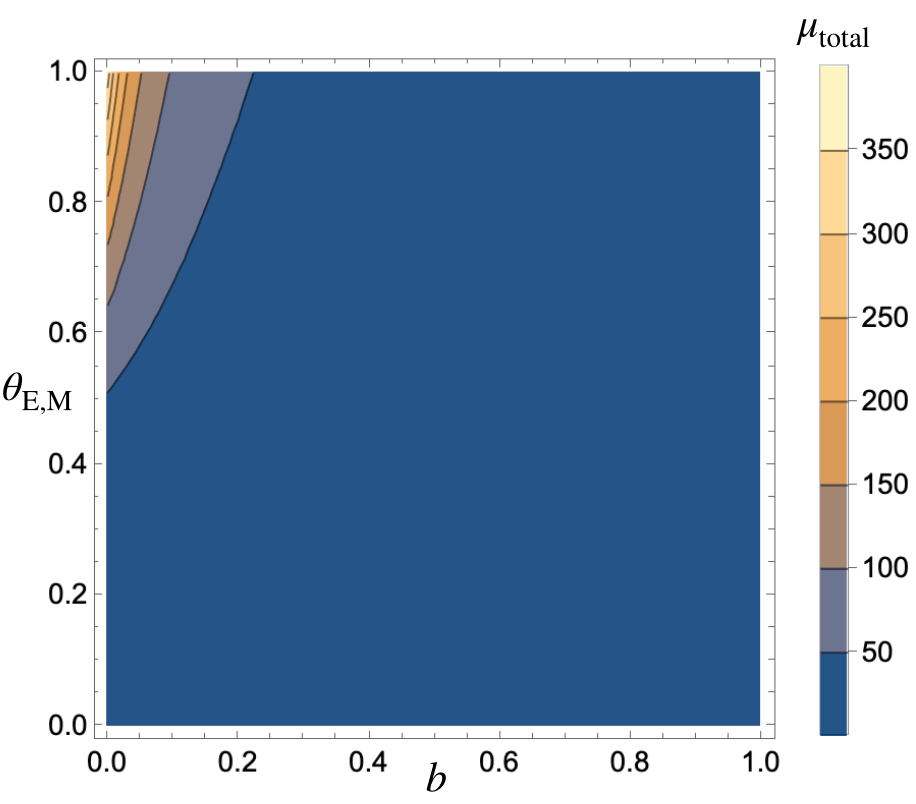}
 	\caption{The contour plot of $\mu_{\rm total}$ in terms of $\theta_{\rm E,Q}$ and $b$ as fixing $\theta_{\rm E,Q}=0.1$. We have set $0<b< 1~\rm kpc$, $D_l=10~\rm kpc$. }
 	\label{mu33}
 \end{figure}

For completeness, we note that $\theta_{\rm E,Q}=0.1$ is a special value since that is almost independent of $\theta_{\rm E,M}$. We numerically obtain that $\mu_{\rm total}$ will be approaching $10^{2}$ at some certain scales of $b$. From figure \ref{mu33}, one could clearly see this feature. 

In this case, we have shown the magnification with three images after bending the light. The range of $\mu_{\rm total}$ is from $1$ to $10^5$. An obvious feature is that the large value of $\mu_{\rm total}$ is mainly determined by $\theta_{\rm E,M}$ included in case \pageref{q=0}. Meanwhile, $\theta_{E,Q}=0.1$ is a very special value that leads to the large value of $\mu_{\rm total}$. In summary, the contradiction between $\theta_{\rm E,M}$ and $\theta_{\rm E,M}$ will lead to the huge change of $\mu_{\rm total}$.

\subsection{Two images }
\label{two and one}
In this part, we will investigate the magnification with two images corresponding to Case 3 of Appendix \pageref{sec:lenssol}. 
In this case, we will uitilize $\Delta=0$ defined by \eqref{delta} that it was the explicit solution between $b$, $\theta_{\rm E,Q}$ and $\theta_{\rm E,M}$. For simplicity, we will introduce variable as $x=\theta_{\rm E,M}^2$ and $y=\theta_{\rm E,Q}^3$, then we can get solution of $x$ in terms of $y$ and $\beta$, 
\begin{equation}
	y=\frac{1}{81} \left(15 \beta  x-2\beta ^3+2\sqrt{\beta ^6-243 x^3+117 \beta ^2 x^2-15 \beta ^4 x})\right),
	\label{y}
\end{equation} 
with the condition $3x< \beta^2$. Next, we will use $\beta\approx \frac{b}{D_l}$ to numerically simulate the magnification. Actually, there are two solutions for $y$. However $y$ should be positive. This is reason that we choose \eqref{y} as the real solution. We should emphasize that  $3x< \beta^2$ implies that $x_{\rm max}=\frac{1}{3}$ since $\beta_{\rm max}=1$. Being armed with these equations, we can numerically simulate the $\mu_{\rm total}$ as follows, 
 \begin{figure}[ht]
 	\includegraphics[scale=0.50]{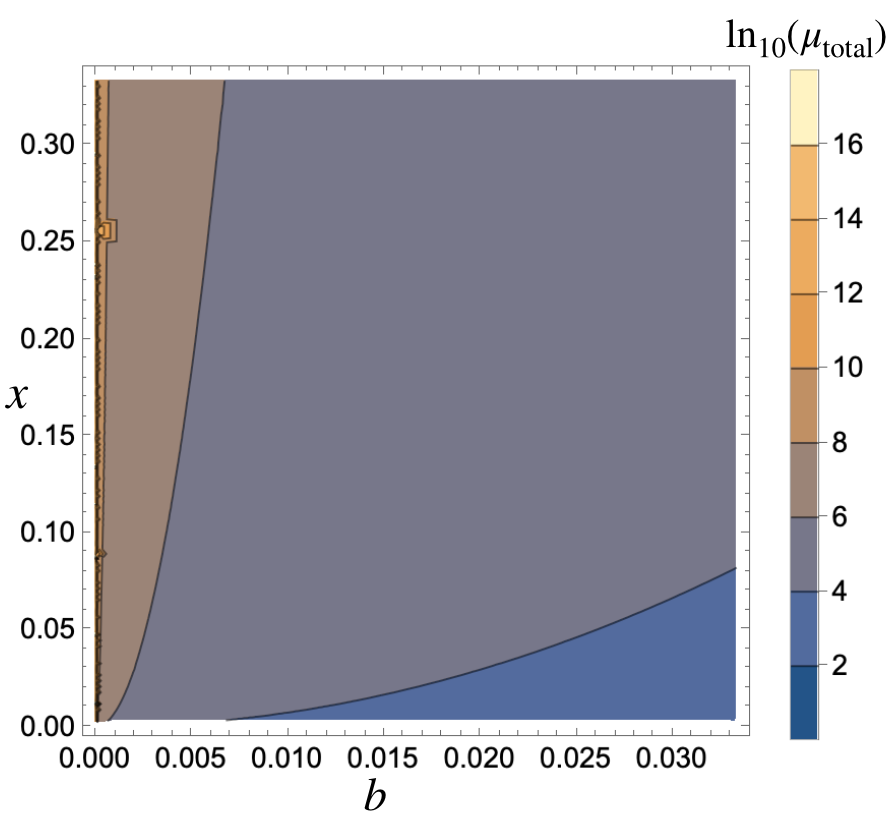}
 	\caption{We have set $0<b<10^{-3}~\rm kpc$ ,$0<x<\frac{1}{3}$ and $D_l=10~\rm kpc $ with $x=\theta_{\rm E,M}^2$. }
 	\label{mu211}
 \end{figure}
Figure \ref{mu211} clearly indicates that $\mu_{\rm total}$ is almost divergent since its magnitude is too large. We also found that $\mu_{\rm total}=2.0$ as $b=10~\rm kpc$ and $x=\frac{1}{3}$ which means that every image will contribute one to the total value of magnification. There is no magnification effect at this scale. 

\subsection{One image}
\label{one}
When there is only one real solution, there are still two cases corresponding to Case 1 and 2 of Appendix VII. For the first one, it is straightforward  for obtaining  $\mu_{\rm total}=\frac{1}{9}$ that is independent of $\theta_{\rm E,Q}$, $\theta_{\rm E,M}$ and $b$. Consequently, this critical value can be dubbed as a criteria for assessing the existence of wormhole. Here, we also list three cases for investigating the total magnification. 

Case a: $\theta_{\rm E,M}\ll 1$ or $\theta_{\rm E,Q}\ll 1 $

In this case, the magnification as the contribution of the mass part is much smaller compared with electric charge part, and vice versa. 
 \begin{figure}[ht]
 	\includegraphics[scale=0.52]{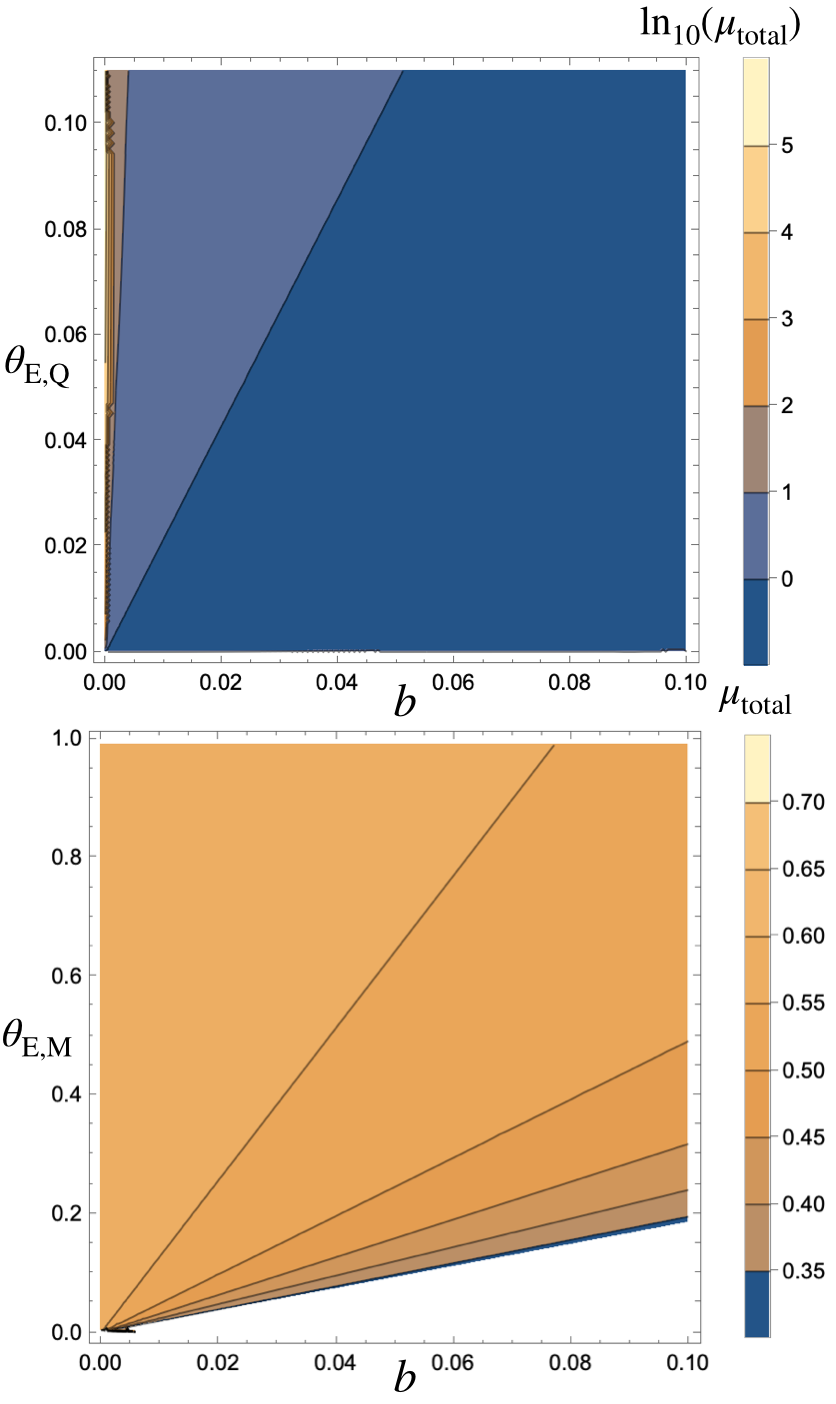}
 	\caption{The upper pannel shows that contour plot of $\ln_{10}(\mu_{\rm total})$, in which we have set $\theta_{\rm E,M}=10^{-4}$ and $0<b<0.1 ~\rm kpc$, $0<\theta_{\rm E,Q}<0.1$. The lower pannel shows the contour plot of $\mu_{\rm total}$ with $\theta_{\rm E,Q}=10^{-4}$ and $0<b<0.1~\rm kpc$, $0<\theta_{\rm E,M}<1$. }
 	\label{mu311}
 \end{figure}
In figure \ref{mu311}, $\mu_{\rm total}$ is of order $10^{5}$ at the very small scale of $b$ and $\theta_{\rm E,Q}$ is varying. As $b$ taking larger values, $\mu_{\rm total}$ approaches unity. 

Next, we will see the other extreme case, where $\theta_{\rm E,Q}\ll 1$ with $\theta_{\rm E,Q}=10^{-4}$.
In the lower pannel of Figure \ref{mu311} reveals a different trend of $\mu_{\rm total}$, in which we will get a smaller image comparing with the light source. And the image will decrease as $b$ is enhanced. 

Case b: $\theta_{\rm E,M} \leq 1$ or $\theta_{\rm E,Q}\leq 1 $

For completeness, we will study the case of $\theta_{\rm E,M} \leq 1$ or $\theta_{\rm E,Q}\leq 1 $.  
 \begin{figure}[ht]
 	\includegraphics[scale=0.52]{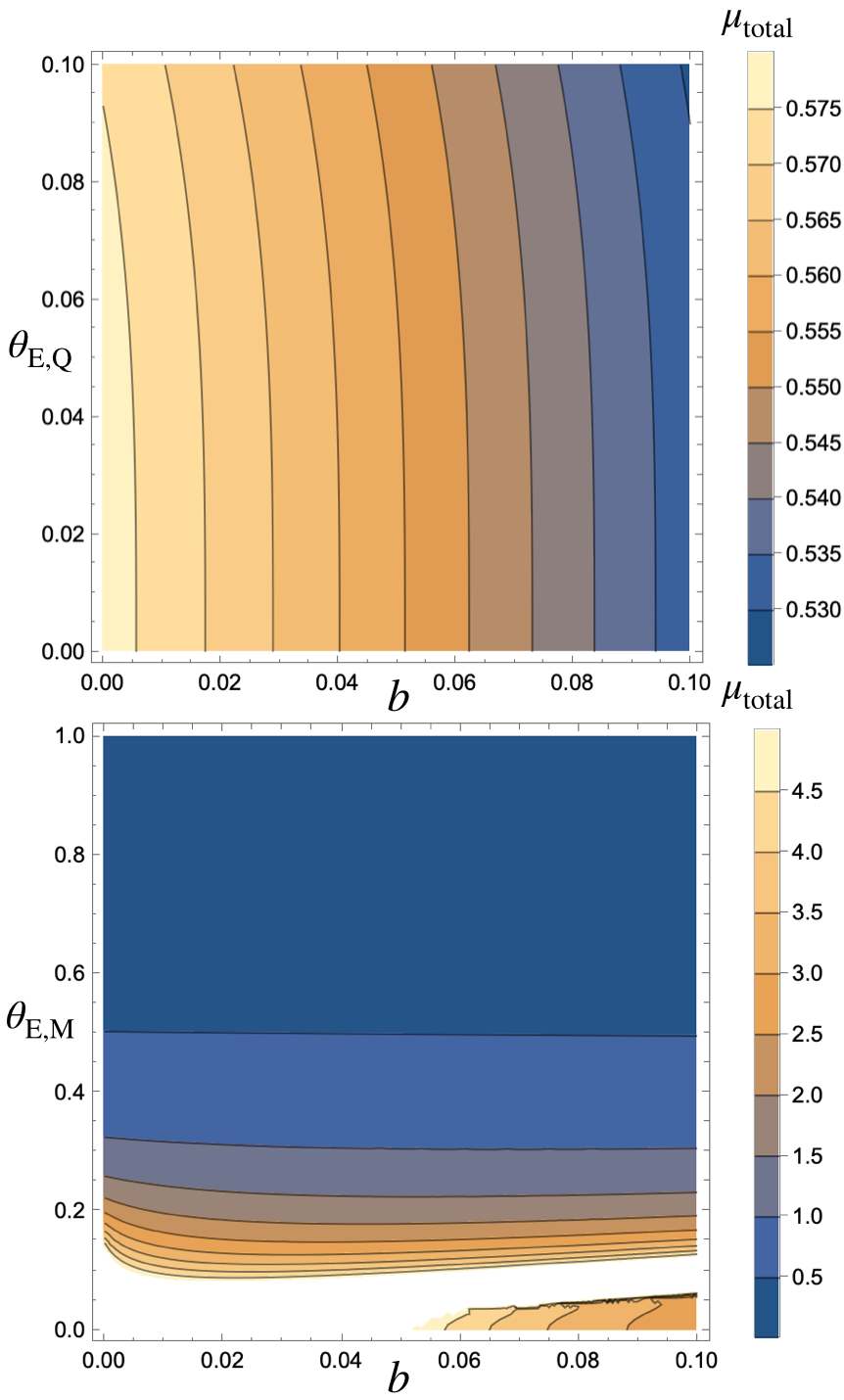}
 	\caption{The upper pannel shows the contour plot of $\mu_{\rm total}$. We have set $\theta_{\rm E,M}=0.8$ and $0<b<0.1~\rm kpc$, $0<\theta_{\rm E,Q}<0.1$. In the lower pannel, contour plot of $\mu_{\rm total}$ with $\theta_{\rm E,Q}=0.8$ and $0<b<0.1~\rm kpc$, $0<\theta_{\rm E,M}<0.11$ }
 	\label{mu333}
 \end{figure}
Figure \ref{mu333} indicates that $\mu_{\rm total}$ is still less than unity. And the $\theta_{\rm E,Q}$ is also set from $0$ to $0.1$, and $\mu_{\rm total}$ varies little in this range.

In the lower pannel of figure \ref{mu333}, we also show that $\mu_{\rm total}$ can reach around $4.5$ at the very small scales of $b$, and we also numerically simulate that $\mu_{\rm total}$ will be of order $10^{11}$ as $b$ and $\theta_{\rm E,M}$ is quite tiny which is covered by the white part of lower pannel of figure \ref{mu333}.

In this section, we have shown the  $\mu_{\rm total}$ with one image varies from $0.5$ to $10^5$, in which the high order of $\mu_{\rm total}$ is different with the case with three images in figure \ref{mu21}. However, we cannot distinguish these two cases if the $\mu_{\rm total}\approx 10^5$ since scale of $b$ is too tiny.

\section{V.~~Conclusion and outlook}
\label{conclusion}
In this paper, we investigate the microlensing effect of a charged wormhole with negative mass. The charge term would contribute a $Q^2/r^2$ term in the effective Newtonian potential, and the resulting lens equation becomes a cubic equation. Hence, in certain situations, the charged wormhole can generate at most three images of a single source. We have systematically investigated the total magnification \eqref{mutotal}, which can be confronted by astrophysical observations. 

According to Appendix VII, we study the magnification with three images \pageref{three images}, two images, and one image \pageref{two and one}: $(a)$ Within three images, our numerical results show that $\mu_{\rm total}$ will be approaching $1.5$ as fixing $\theta_{\rm total}=10^{-7}$ dubbed as a criterion for assessing there are three images in figure \ref{mu11}. If there was a peak around $100$, it can also be explained by figure \ref{mu12}. Figure \ref{mu21} indicates that $\mu_{\rm total}$ could reach the order $10^{6}$ as setting $\theta_{\rm E,Q}\ll 1$. And we also found that the $\mu_{\rm total}$ will be of the order of $10^{3}$ as enhancing the value of $\theta_{\rm E,Q}$ when the mass and charge part become compatible. (b) For the two images after bending the light, $\mu_{\rm total}$ will be divergent at the very small scales, and $\mu_{\rm total}$ will be approaching $2$ as $b>0.1~\rm kpc$. (c) The last case with only one image, we could see that $\mu_{\rm total}=\frac{1}{9}$ that is independent of $b$, $\theta_{\rm E,Q}$ and $\theta_{\rm E,M}$ corresponding to $\Delta_1=\Delta_2=0$. Another one is that $\mu_{\rm total}$ will also be quite large as $\theta_{E,M}=10^{-4}$ that is opposite with figure \ref{mu21}. The other figures show that the range of $\mu_{\rm total }$ is from $10^{-2}$ to $10$. As for $\mu_{\rm total}=10^5$, we cannot distinguish the case that has three images or only one image.  At last, the most important thing is that  there will be two maximal values of magnification (one is peak, the other one is gentle peak) when the contribution via mass is much less than the charge part, for which there are three images after bending the light ray, which is of significance for exploring the origin of corresponding observations.

Our work is a preliminary investigation on this topic, and there are many interesting topics to explore in the future. Firstly, we in this work only evaluate the magnification of the lensing effect, and other important parameters such as the event rate are not included. To test a more complete analysis is required. Besides, the current work assumes the source to be far from the wormhole throat, so that the wormhole can be simply treated as a point mass with charge. Although this treatment is generic for any charged wormhole, we lose the information about the microscopic physics near the wormhole's throat. Hence, we can explore the lensing property when the source is near the wormhole throat with different charged wormhole models. By confronting the results with future observations, we may get a chance to explore fundamental physics through microlensing. 

Finally, the lens equations may be similar in different physical systems. For example, \cite{Bozza:2015wbw, Bozza:2020ubm} studied the lensing effect of a binary system, and the lensing by charged black holes are investigated in \cite{Horvath:2010xq,Horvath:2012ru,Zhao:2016kft,Wang:2019cuf,Kumar:2020sag,Neves:2020doc,Javed:2021arr}. These works share similar but not equivalent lens equations with us. It would be important to investigate if these models can be distinguished through astrophysical observations. Even we could test the emergent gravity by its deflection angle \cite{Liu:2016nwt}. Based on the lensing effect, we can extend \cite{Chakraborty:2016lxo} into the framework of microlensing to explore the extra dimension.

 \section*{Acknowledgements}
 We appreciate the stimulating discussions with Bichu Li, Yuhang Zhu, Chao Chen, and Ke Gao. 
 LH is funded by NSFC grant NO. 12165009. This work is supported in part by the National Key RD Program of China (2021YFC2203100). YFC is supported in part by the NSFC (Nos. 11961131007, 11653002), the National Youth Talents Program of China, the Fundamental Research Funds for Central Universities, the CSC Innovation Talent Funds, by the CAS project for Young Scientists in Basic Research (YSBR-006), and by the USTC Fellowship for International Cooperation. WL acknowledge the support from the National Key RD Program of China (2021YFC2203100), NSFC(NO.11833005, 12192224). MZ and YW are supported in part by the CRF grant C6017-20GF, the GRF grant 16303621 by the RGC of Hong Kong SAR, and the NSFC Excellent Young Scientist Scheme (Hong Kong and Macau) Grant No. 12022516.

 \appendix
 \section*{A. Appendix:~The solutions of lensing eq. \eqref{eq:lensbeta}}
\label{sec:lenssol}
In this Appendix, we present the solution to the lens equation \eqref{eq:lensbeta}:
\begin{align}
\beta = \theta + \frac{\theta_{E,M}^2}{\theta} - \frac{\theta_{E,Q}^3}{\theta^2} ~\to~ \theta^3 - \beta \theta^2 + \theta_{E,M}^2 \theta - \theta_{E,Q}^3 = 0 ~.
\end{align}

Generically, there are three complex solutions to \eqref{eq:lensbeta}. For convenience, we define the following discriminant
\begin{align}
&\Delta_1 = \beta^2 - 3\theta_{E,M}^2 ,\\& \Delta_2 = 9\theta_{E,Q}^3 - \beta \theta_{E,M}^2 ~,
\\& \Delta_3 = \theta_{E,M}^4 - \beta \theta_{E,Q}^3 ~,
\end{align}
and an overall discriminant
\begin{equation}
\Delta = \Delta_2^2 - 4 \Delta_1 \Delta_3 ~,
\label{delta}
\end{equation}
then the discriminant decide the number of real solutions:

\begin{enumerate}
	\item $\Delta_1 = \Delta_2 = 0$, then there is one real solution $\theta = \frac{\beta}{3}$.
	
	\item $\Delta > 0$, then there is one real solution, 
	\begin{align}
	&\theta = \frac{1}{3} \bigg[ \beta - \left( -\beta \Delta_1 - \frac{3}{2} (\Delta_2 - \sqrt{\Delta})  \right)^\frac{1}{3}\nonumber \\& - \left( -\beta \Delta_1 - \frac{3}{2} (\Delta_2 + \sqrt{\Delta})  \right)^\frac{1}{3}  \bigg] ~.
	\label{theta2}
	\end{align}
	
	\item $\Delta = 0$, then there are two real solutions
	\begin{equation}
	\theta_1 = \beta + \frac{\Delta_2}{\Delta_1} ~,~ \theta_2 = -\frac{\Delta_2}{2\Delta_1} ~.
	\end{equation}
	
	\item $\Delta < 0$, then there are three real solutions for $\Delta_1 > 0$ case:
	\begin{align}
	&\theta_1 = \frac{\beta}{3} - \frac{2}{3} \sqrt{\Delta_1} \cos \frac{\varphi}{3} ,\\
	& \theta_{2,3} = \frac{\beta}{3} + \frac{1}{3} \sqrt{\Delta_1} \left( \cos \frac{\varphi}{3} \pm \sqrt{3} \sin \frac{\varphi}{3} \right) ~,
	\end{align}
	where 
	\begin{equation}
	\cos \varphi = - \frac{2\beta \Delta_1 - 3 \Delta_2}{2\Delta_1^{\frac{3}{2}}} ~,
	\end{equation}
	is an auxiliary variable. When $\Delta_1 < 0$, the three solutions are all complex.
	
\end{enumerate}

We see that the property of images are highly dependent on $\Delta$. Moreover, the parameters $\theta_{E,M}$ and $\theta_{E,Q}$ are determined by the property of a wormhole and the angular distance, so they are fixed when the source is moving nearby the wormhole. Hence, we are interested in the positivity of $\Delta$ when $\beta$ is varying. Organize $\Delta$ as a function of $\beta$, we have
\begin{equation}
\label{eq:Delta}
\Delta = 4\theta_{E,Q}^3 \beta^3 - 3\theta_{E,M}^4 \beta^2 - 30 \theta_{E,M}^2 \theta_{E,Q}^3 \beta + 12 \theta_{E,M}^6 + 81 \theta_{E,Q}^6 ~.
\end{equation}

Notice that, when $\theta_{E,Q} = 0$, the expression \eqref{eq:Delta} becomes $\Delta = -3\theta_{E,M}^4 (\beta^2 - 4\theta_{E,M}^4)$, and we recover the criticle Einstein angle $\theta_{E,M} = \frac{1}{2}\beta$ for a wormhole with negative mass.

 \section*{B. Appendix:~The derivation of effective Newtonian potential from isotropic coordinate}
\label{app b}

In this appendix, we will prove that the Newtonian potential is the same under the isotropic coordinate. The isotropic coordinate is defined such that the light cones appear round, and the spatial part of the metric is conformally flat. Its metric reads
\begin{equation}
	ds^2=-A^2c^2dt^2+B^2\big[d\rho^2+\rho^2(d\theta^2+\sin^2 \theta d\phi^2)\big],
	\label{generic iso}
\end{equation}
where $0<\theta<\pi$ and $0<\phi<2\pi$ are the conventional angular coordinate. In static case, the functions $A \equiv A(\rho)$ and $B \equiv B(\rho)$ depends only on the radial coordinate $\rho$. The isotropic metric \eqref{ellis huang1} is related to our metric \eqref{generic iso} by requiring
\begin{equation}
	\label{eq:coordinatetrans}
	A^2 = h ~,~ B^2 \left( \frac{d\rho}{dr} \right)^2 = \frac{1}{h} ~,~ B^2\rho^2 = r^2 ~,
\end{equation}
when $\sigma \simeq 1$.

Thus, the coordinates are related by
\begin{equation}
	\left( \frac{d\rho}{dr} \right)^2 = \frac{\rho^2}{r^2} \frac{1}{h(r)} ~.
\end{equation}
Take the positive branch, we have
\begin{equation}
	\label{eq:drdrho}
	\frac{d\rho}{\rho} = \frac{dr}{r\sqrt{h(r)}} = \frac{dr}{\sqrt{r^2 + h_M M r + h_Q Q_e^2}} ~.
\end{equation}
and
\begin{equation}
	\label{eq:rhoasr}
	4\rho = 2r + h_MM + 2\sqrt{r^2 + h_M Mr + h_Q Q_e^2} ~.
\end{equation}
Now $r$ depends on $\rho$ as
\begin{equation}
	\label{eq:rasrho}
	r = \rho \left( 1 - \frac{h_M M}{4\rho} \right)^2 - \frac{1}{4\rho} h_Q Q_e^2 ~.
\end{equation}

Note that in \eqref{eq:rhoasr}, the left hand side should be $C \rho$ with $C$ an integration constant. The constant $C$ corresponds to a coordinate redefinition $\rho \to C\rho$, so we can take any value in \eqref{eq:rhoasr} without loss of generality. Here we take $C=4$, so that $r \simeq \rho$ when $M \ll 1$ and $Q_e \ll 1$. Moreover, when $h_M = -2$ and $h_Q = 0$ we recover the relationship for Schwarzschild metric
\begin{equation}
	r = \rho \left( 1 + \frac{M}{2\rho} \right)^2 ~.
\end{equation}

With the help of \eqref{eq:coordinatetrans} and \eqref{eq:rasrho}, the expressions of functions $A$ and $B$ are direct:
\begin{equation}
	B(\rho) = \frac{r}{\rho} = \left( 1 - \frac{h_M M}{4\rho} \right)^2 - \frac{1}{4\rho^2} h_Q Q_e^2 ~,
\end{equation}
\begin{equation}
	A(\rho) = \sqrt{h} = \left[ 1 + \frac{h_M M}{\rho B(\rho)} + \frac{h_Q Q_e^2}{\rho^2B^2(\rho)}  \right]^{\frac{1}{2}} ~.
	\label{Arho}
\end{equation}

Finally, to recover the conventional expressions in the weak gravity approximation, we note that $h = \mathcal{O}(1)$, and the deviation $h - 1 = \frac{h_M M}{r} + \frac{h_Q Q_e^2}{r^2}$ should be treated as leading order. Moreover, we have $r/\rho \simeq 1$, so any term with $\left( \frac{h_M M}{\rho} \right)^2$, $\left( \frac{h_M M}{\rho} \right) \left( \frac{h_Q Q_e^2}{\rho^2} \right)$ or $\left( \frac{h_Q Q_e^2}{\rho^2} \right)^2$ are of secondary order. Then we write $B(\rho)$ in a more familiar form
\begin{equation}
	B(\rho) \simeq \left( 1 - \frac{h_M M}{4\rho} - \frac{h_Q Q_e^2}{8\rho^2} \right)^2 ~.
	\label{Brho}
\end{equation}
Similarliy, we have
\begin{equation}
	AB \simeq \left( 1 - \frac{h_M^2 M^2}{8\rho^2} - \frac{h_Mh_Q MQ_e^2}{8\rho^3} + \frac{3h_Q^2 Q_e^4}{32\rho^4} \right)^{\frac{1}{2}} ~. 
\end{equation}
Since the right hand side differs from unity only at secondary order, at weak gravity regime we have
\begin{align}
	& \nonumber ds^2 = -c^2 \left( 1 + \frac{h_M M}{4\rho} + \frac{h_Q Q_e^2}{8\rho^2} \right)^4 dt^2 \\
	& + \left( 1 - \frac{h_M M}{4\rho} - \frac{h_Q Q_e^2}{8\rho^2} \right)^4 \left( d\rho^2+\rho^2d\Omega^2 \right) ~.
\end{align}
When we get the first order from this metric, one can straightforwardly obtain the Newtonian potential under the isotropic coordinate as follows, 
\begin{equation}
		\Phi(\rho)=-\frac{h_MMc^2}{2\rho}-\frac{h_QQ_e^2c^2}{2\rho^2},
		\label{isotropic newton potential}
\end{equation}
which is in accordance with Eq. \eqref{eq:Phi}. Finally, we need to clarify one issue which is the correction to this potential. Actually, one can explicitly obtain the exact the potential from Eq. \eqref{Arho} and \eqref{Brho}. However, what we are paraticularly interested in the weak field region where the Newtonian potential \eqref{eq:Phi} is sufficient for microlensing effects.

\end{document}